\documentclass[letterpaper, 10 pt, conference]{ieeeconf}

\IEEEoverridecommandlockouts 
\usepackage{amsmath,amssymb,amsfonts,enumerate, mathtools,mathrsfs}
\usepackage{graphics}
\usepackage{epsfig}
\usepackage{times}
\usepackage{booktabs}
\usepackage{multirow}
\usepackage{array}
\usepackage{algpseudocode}
\usepackage[colorlinks=true,allcolors=black]{hyperref}
\usepackage{xcolor}
\usepackage{bm}
\usepackage{cite}
\let\labelindent\relax
\usepackage{enumitem}
\usepackage{subfigure}
\usepackage{diagbox}
\usepackage{caption}
\usepackage{dblfloatfix}
\usepackage{esvect}

\usepackage[linesnumbered,ruled,vlined]{algorithm2e}
\usepackage{url}

\newtheorem{defn}{Definition}

\newcommand{\tool}{\textsc{Feat2Map}\xspace}

\begin{document}

\title{\LARGE \bf Automatic Map Generation for Autonomous Driving System Testing}

\author{Yun Tang, Yuan Zhou, Kairui Yang, Ziyuan Zhong, Baishakhi Ray, Yang Liu, Ping Zhang, Junbo Chen
\thanks{Yun Tang, Yuan Zhou and Yang Liu are with Nanyang Technological University, Singapore 639798. Email: \tt\small yun005@e.ntu.edu.sg, \tt\small \{y.zhou, yangliu\}@ntu.edu.sg}
\thanks {Kairui Yang, Ping Zhang and Junbo Chen are with DAMO Academy, Alibaba Group, China. Email: \tt\small kairui.ykr@alibaba-inc.com, \tt\small zp.zp@cainiao.com, \tt\small junbo.chenjb@taobao.com 
}
\thanks {Ziyuan Zhong and Baishakhi Ray are with Columbia University, New York, NY, 10025. Email: \tt\small ziyuan.zhong@columbia.edu, \tt \small rayb@cs.columbia.edu}
}

\maketitle
\begin{abstract}

High-definition (HD) maps are essential in testing autonomous driving systems (ADSs). HD maps essentially determine the potential diversity of the testing scenarios. However, the current HD maps suffer from two main limitations: lack of junction diversity in the publicly available HD maps and cost-consuming to build a new HD map.  Hence, in this paper, we propose the first method, \tool, to automatically generate concise HD maps with scenario diversity guarantees. \tool focuses on junctions as they significantly influence scenario diversity, especially in urban road networks. \tool first defines a set of features to characterize junctions. Then, \tool extracts and samples concrete junction features from a list of input HD maps or user-defined requirements. Each junction feature generates a junction. Finally, \tool builds a map by connecting the junctions in a grid layout. To demonstrate the effectiveness of \tool, we conduct experiments with the public HD maps from SVL and the open-source ADS Apollo. The results show that \tool can (1) generate new maps of reduced size while maintaining scenario diversity in terms of the code coverage and motion states of the ADS under test, and (2) generate new maps of increased scenario diversity by merging intersection features from multiple maps or taking user inputs.


\end{abstract}


\section{Introduction}

Autonomous driving systems (ADSs) have gained enormous attention and development recently. Safety is the public's primary concern for the mass deployment of ADSs.
According to ISO/PAS 21448 \cite{iso2019pas}, SOTIF (Safety Of The Intended Functionality) is the key for ADS safety. 
However, due to the enormous number of parameter combinations to specify test scenarios, it is still a challenge to validate the SOTIF of ADSs, which aims to reduce the region of unknown and unsafe scenarios. 

Among the existing methods for the validation of SOTIF, simulation-based ADS testing is a major approach to discovering unknown and unsafe scenarios for ADSs~\cite{zhong2021survey}.
In the simulation environment, we can model different elements for an autonomous vehicle, such as HD (High-Definition) maps (e.g., road network, traffic signs, and traffic lights), vehicle dynamics, and sensors (e.g., camera, LiDAR, radar).
Among all the elements, HD maps are prerequisites for ADS testing, and they essentially determine the potential diversity of the testing scenarios, directly impacting the testing efficiency. For example, Gambi \emph{et al.} demonstrated that many lane departure failures have manifested on different shapes of roads~\cite{gambi2019automatically}. 
Tang \emph{et al.} showed that by exploiting the topology features of junctions in HD maps, diverse testing scenarios could be generated, and different issues related to SOTIF were exposed in the Apollo ADS~\cite{tang2021route,tang2021collision,tang2021systematic}. 

Despite the importance of HD maps, there are some limitations of the publicly available HD maps during ADS testing.
First, the commercial HD maps, such as those provided by \cite{web_tomtom, web_here} are replicas of the real world and contain a significant amount of duplicate map elements (e.g., junctions with similar shapes).
Thus, scenarios from such HD maps may lack diversity.
Second, although the open-source simulators, such as Carla \cite{carla_simulator} and SVL \cite{svl_simulator}, often come with some built-in HD maps, they are relatively small with a limited number of junctions and roads. 
Consequently, the diversity of scenarios generated from these maps is limited. 
Third, different cities or regions may contain different junction structures. Hence, to test ADSs sufficiently, we need different HD maps from different areas and perform ADS testing within each map. It is time-consuming.
Fourth, manually building diverse HD maps using tools, such as SUMO netedit \cite{web_sumo_netedit} or CommonRoad Scenario Editor \cite{web_commonRoad_editor}, is laborious, given a large number of scenario requirements. 
Therefore, it is essential to develop a framework to generate concise but scenario-diverse HD maps automatically. 

However, there is little work on automatic HD map generation. Gambi \emph{et al.} \cite{gambi2019automatically} proposed a search-based approach to generate lanes of changing curvatures to test the lane-keeping functionality. However, other traffic elements, such as intersections and traffic controls, are not considered. Thus, the map generated is incomplete, and the scenarios are limited. Mi \emph{et al.} \cite{mi2021hdmapgen} used a trained hierarchical graph generation model to generate the road network graphs. The model is trained with existing HD maps and tries to generate maps similar to its training data. As a result, it suffers from the same aforementioned limitations (e.g., limited scenario diversity and duplicated intersections) as the training data (i.e., existing HD maps).

In this paper, we propose the first feature-based method, \tool, to generate scenario-diverse HD maps automatically based on junction features as junctions significantly affect scenario diversity, especially in urban road networks \cite{tang2021route,tang2021collision,tang2021systematic}.
First, we define a feature vector to characterize junctions, including the number of roads connected by a junction, orientations of the connected roads, traffic controls, and crosswalk features.
Second, \tool extracts all possible values of each feature element from a set of maps or users' inputs.
Third, \tool applies combinatorial sampling to generate a minimal set of concrete feature vectors without duplication while covering all possible feature combinations. 
Finally, \tool builds a junction per concrete feature vector and constructs a complete HD map with connected junctions in a grid layout. 
To demonstrate the effectiveness of \tool, we generated several HD maps from single and multiple input maps and performed qualitative and quantitative analysis by performing route coverage testing \cite{tang2021route} on the latest open-source ADS Apollo 7.0 \cite{web_apollo_github}. 
The results show that (1) \tool can effectively extract features from an input map and generate concise HD maps while preserving scenario diversity; (2) \tool can generate HD maps based on combined features from multiple input maps with higher perceived scenario diversity by the ADS-under-test than individual input map; and (3) \tool can generate customized HD maps from user input.



\section{Preliminaries and Problem Statement}\label{problem}
\label{section:problem}


\begin{figure}
    \centering
    \includegraphics[width=\linewidth]{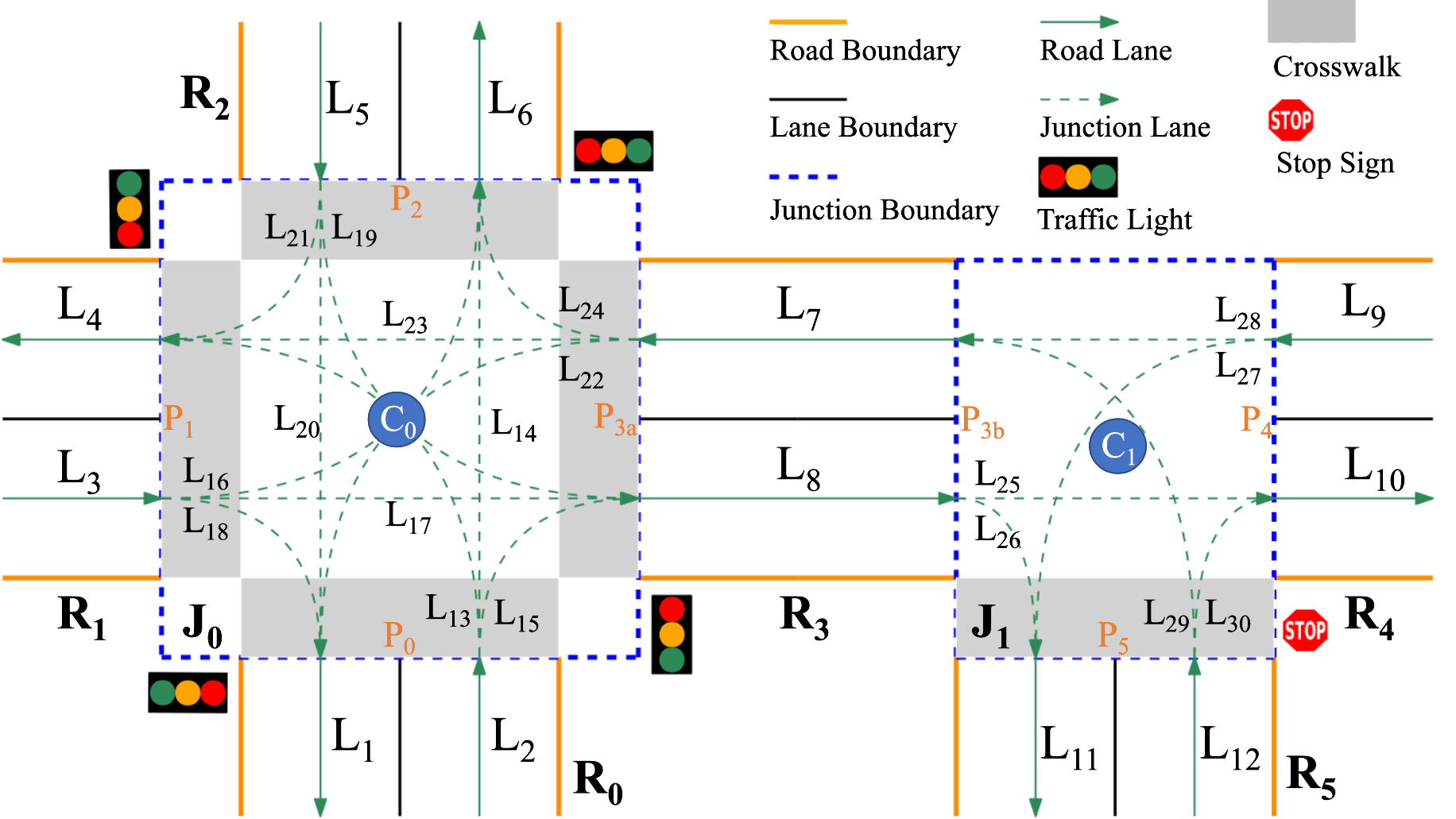}
    \caption{A two-junction traffic network.}
    \label{fig:cross} 
\end{figure}

HD map is one of the essential components for modern ADSs, especially in city-driving scenarios. It encodes the semantic features (e.g., topology and geometry) of traffic network elements such as traffic lanes, roads, intersections, traffic lights, crosswalks, and stop signs~\cite{web_HDmap}. Fig. \ref{fig:cross} shows a diagram of an HD map containing a two-junction traffic network. There are six two-way roads ($R_0-R_5$) in the diagram. Each road (e.g., $R_0$) contains a forward lane (e.g., $L_2$ in $R_0$) and a backward lane (e.g., $L_1$ in $R_0$). $R_0-R_3$ form a traffic light-controlled junction $J_0$, and $R_3-R_5$ construct a stop sign-controlled junction $J_1$. 
Lanes are the atomic element in an HD map, and each lane is represented by its center reference line. 
Lanes in an HD map can be classified into road lanes (e.g., $L_1-L_{12}$ in Fig. \ref{fig:cross}) and junction lanes (e.g., $L_{13}-L_{30}$ in Fig. \ref{fig:cross}).

Given an HD map, ADS testing aims to guarantee that the autonomous vehicle can run safely at any part of the map under any scenario. 
However, an HD map from part of a city or a limited area usually contains limited intersection structures.
Consequently, we cannot test an ADS sufficiently on such a map. 
Hence, in this paper, we aim at \emph{generating an HD map automatically by eliminating duplicated intersections
while maximizing the diversity of intersection structures and potential scenarios.} 

Before illustrating the generation method in detail, we first discuss some assumptions in the paper. 
(1) We currently focus on the intersections of a map, as intersections can significantly affect the map complexity and scenarios diversity. In the future, we will take roads into consideration.
(2) An ADS can be roughly divided into the perception and the planning components. We mainly focus on generating maps for testing the planning component. Testing of AI models in the perception component is orthogonal to our work.
\section{Methodology}


\subsection{Modeling of Lanes}
\label{section:road_model}
Based on \cite{web_openDRIVE}, lanes can be represented by their center reference paths. We model these reference paths using cubic Bézier curves, which are extensively used in computer graphics~\cite{web_bezier_curve_wiki}. 
Given four control points $P_0$, $P_1$, $P_2$ and $P_3$, the cubic Bézier curves can be formulated as: 
\begin{equation}
    B(t) = \sum_{i=0}^3 \binom{3}{i} (1-t)^{3-i} t^i P_i
\end{equation}
where $t\in[0,1]$, and $\binom{3}{i}$ are the binomial coefficients. 
Cubic Bézier curves are powerful in representing simple curves such as straight lines, circle-shaped or S-shaped, and can be combined into Bézier splines of complex shapes. 
For any path (e.g. reference path or boundary curve of a lane), if the \emph{start point} $P_{start}$, \emph{start heading} $\theta_0$, \emph{end point} $P_{end}$, and \emph{end heading} $\theta_1$ are known, the controls points can be calculated as: 
\begin{equation}
    \begin{split}
        P_0 & = P_{start} \\
        P_1 & = P_{start} + (d_0\cos\theta_0,  d_0\sin\theta_0) \\
        P_2 & = P_{end} - (d_1\cos\theta_1, d_1\sin\theta_1) \\
        P_3 & = P_{end}
    \end{split}
\end{equation}
where both $d_0$ and $d_1$ are heuristically defined as half of the distance between $P_0$ and $P_3$ for a smoother curve.

\subsection{Feature Modeling of Junctions}
\label{section:junction_model}
In the section, we define the \emph{junction features} to characterize a junction, containing the road feature, the rotation feature, the control feature, and the crosswalk feature. 

First, the \emph{road feature} aims to describe the number of roads connected by a junction, such as T-junctions and n-legged intersections.
The more connected roads, the more complex the junction becomes, resulting in more testing efforts for ADSs. 
Hence, the road feature of a junction $J_j$, denoted as $F_{J_j}^{road}$, is defined as the number of roads connected by a junction.
For example, the junction $J_0$ shown in Fig. \ref{fig:cross} connects four roads, so $F_{J_0}^{road} = 4$, while $F_{J_1}^{road} = 3$.

Second, the \emph{rotation feature} aims to model the relative orientations of the roads connected by a junction.
It indicates the relative directions an autonomous vehicle can drive after entering the junction.
Given a junction $J_j$, suppose its center is $C$, which is calculated as the geometry center of its bounding polygon, and the corresponding start point of road $R_i$ is $P_i$, which is the connecting endpoint of $R_i$’s reference line to the junction.
The vector $\vv{CP_i}$ is called \emph{road socket} of $R_i$.
Hence, the relative location of $R_i$ and $R_k$ can be measured by the angle between $\vv{CP_i}$ and $\vv{CP_k}$.
Hence, the rotation feature is a vector of angles, each describing the angle from one road socket to its adjacent one in counter-clockwise order. 
Given a junction $F_j$, its rotation feature is denoted as $F_{J_j}^{rot}=[\gamma_1, \gamma_2, \ldots, \gamma_N]$, where $N=F_{J_j}^{road}$ and $\sum_{i=0}^N \gamma_i = 2\pi$.

\begin{figure}
    \centering
    \includegraphics[width=0.8\columnwidth]{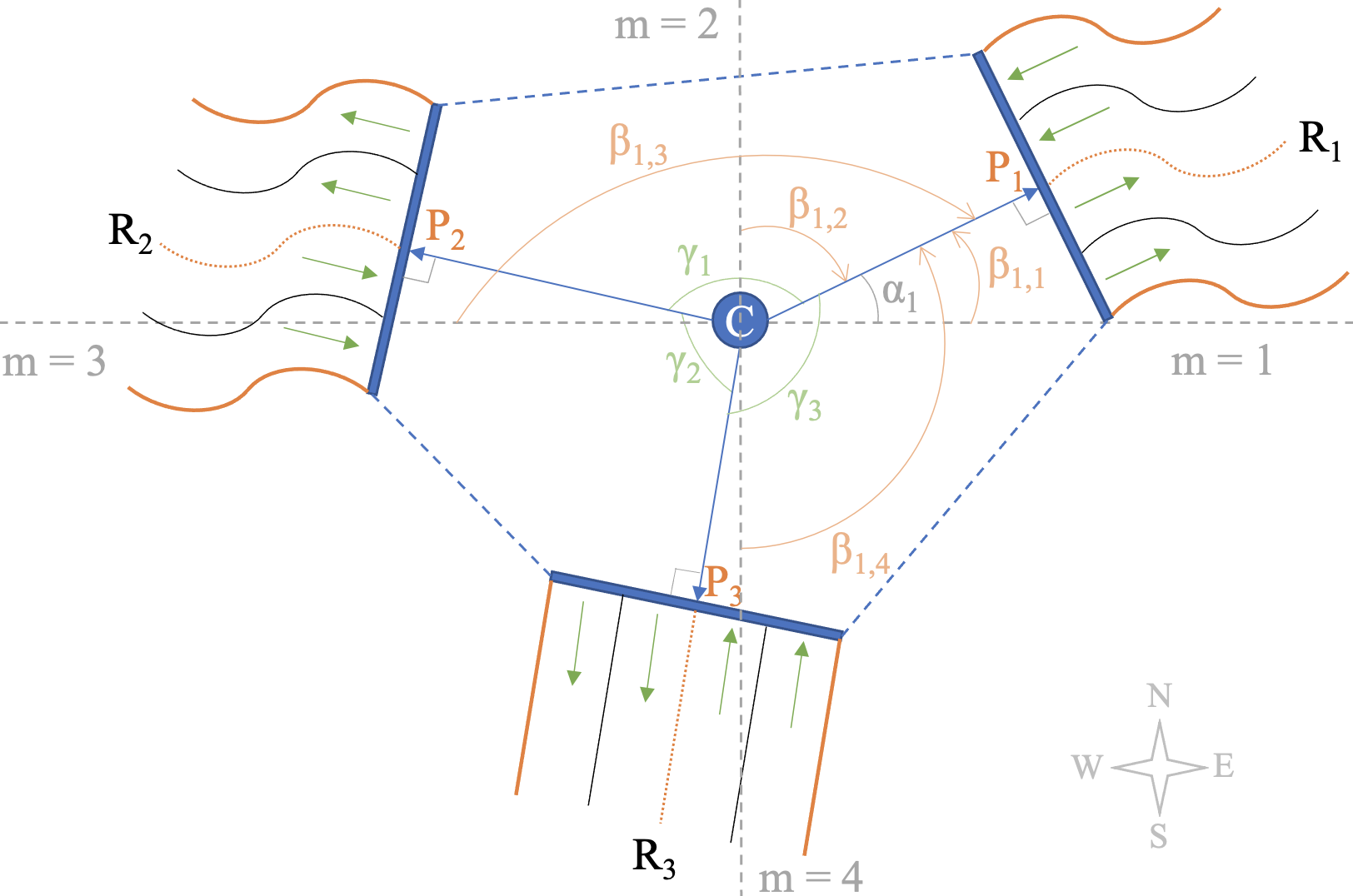}
    \caption{Junction rotation normalization.}
    \label{fig:junction_model}
\end{figure}

Given such a rotation feature, we can generate various junctions by rotating the original junction around its center. 
Hence, we may first perform rotation normalization, i.e., rotating each junction such that it aligns to the four regular directions, i.e., \texttt{East}, \texttt{North}, \texttt{West}, and \texttt{South}, as much as possible.
The rotation normalization process can be described as follows. 
Suppose the rotation feature of a junction $J_j$ is $[\gamma_1, \gamma_2, \ldots, \gamma_N]$, where $N = F_{J_j}^{road}$. 
We first rotate and re-index the road sockets and the angles such that $\gamma_1$ is the maximal one. 
After rotation normalization, the angle from the \texttt{East} direction to $CP_1$ is $\alpha_{opt}$.
Then, $\alpha_{opt}$ can be the optimal solution by solving the following optimization problem:
\begin{align}
    &\min_{\alpha} \sum_{i=1}^N \sum_{m=1}^4 sign_{i,m} \beta_{i,m}^2  \\
    \text{s.t. }  \forall m & \in \{1,2,3,4\}, i\in \{2,\ldots, N\},\\
     & sign_{i,m} \in \{0, 1\},\\
    & \beta_{1,m} = \alpha - (m-1)\pi/2, \label{eqn:angle1}\\
    & \beta_{i,m} = \alpha + \sum_{k=1}^{i-1} \gamma_{k} - (m-1)\pi/2,\label{eqn:angle2}\\
    & \sum_{m=1}^4 sign_{i,m} = 1.
\end{align}
were $\beta_{i,1}$ (resp., $\beta_{i,2}$, $\beta_{i,3}$, and $\beta_{i,4}$) is the angle from \texttt{East} (resp., \texttt{North},  \texttt{West},  \texttt{South}) to $\vv{CP_i}$.
Thus, the rotation feature can be represented by the angles from the \texttt{East} direction to the road sockets, i.e., $F_{J_j}^{rot} =[\alpha_1, \alpha_2, \ldots, \alpha_N]$,  where $\alpha_i=\beta_{i,1}$ is computed by Eqns. \ref{eqn:angle1} and \ref{eqn:angle2} with $\alpha=\alpha^{opt}$.
With the normalized rotation feature, the corresponding road sockets are uniquely determined.
For example, as shown in Fig. \ref{fig:junction_model}, suppose after rotation normalization,  $\alpha^{opt}=\alpha_1$, then we have the rotation feature $F_{J_j}^{rot} = [\alpha_1, \alpha_1 + \gamma_1, \alpha_1 + \gamma_1 + \gamma_2]$.

Third, the \emph{control feature} designates how the traffic is regulated in a junction.
Commonly, there exist three kinds of control manners, i.e., $bare$: no controls, $signal$: controlled by traffic lights, and $stop$: controlled by stop signs.
Hence, the control feature can be defined as $F_{J_j}^{ctrl} \in \{bare, signal, stop\}$. 
For example, in Fig. \ref{fig:cross}, $F_{J_0}^{ctrl} = signal$ and $F_{J_1}^{ctrl} = stop$.

Finally, the \emph{crosswalk feature}, denoted as $F_{J_j}^{xwlk}$, indicates whether there are crosswalks in a junction. Hence, $F_{J_j}^{xwlk}\in \{True, False\}$.

Therefore, we have the following definition. 
\begin{defn}\label{defn:feature}
The junction feature of a junction $J_j$, denoted as $\mathcal{F}_{J_j}$, is a vector defined as:
\begin{equation}
    \mathcal{F}_{J_j} = [ F^{road}_{J_j}, F^{rot}_{J_j}, F^{ctrl}_{J_j}, F^{xwlk}_{J_j}],
\end{equation} 
where $F^{road}_{J_j}\in \mathbb{N}_{\geq 3}=\{3,4,\ldots\}$ and $F^{rot}_{J_j} \in \prod_{i=1}^{F^{road}_{J_j}} [0, 2\pi)$ is a vector of length $F^{road}_{J_j}$.
\end{defn}

\subsection{Extraction of Feature Vectors}
According to Definition \ref{defn:feature}, we describe how to extract the features from a set of HD maps.

First, since the road, the control, and the crosswalk features are discrete variables, we can obtain all their values by checking the junctions in each map.
Therefore, the sets of all possible values of the three features of a given map are denoted as $\mathcal{F}^{road}$, $\mathcal{F}^{ctrl}$, and $\mathcal{F}^{xwlk}$, respectively.

Second, as the rotation feature contains continuous variables, we collect from the HD maps the minimal and the maximal values of each element in the rotation feature. 
Note that the rotation features from the junctions with the same road feature can be compared element-wisely after rotation normalization.
Hence, the collected rotation feature for the junctions with $n$ connected roads can be described as:
\begin{equation}
    \mathcal{F}_n^{rot} = [[\alpha_1^{\min}, \alpha_1^{\max}], \ldots, [\alpha_n^{\min}, \alpha_n^{\max}]],
\end{equation}
where $\alpha_i^{\min} = \min\limits_{J_j\in \mathcal{J}^n} F_{J_j}^{rot}[i]$, 
$\alpha_i^{\max} = \max\limits_{J_j\in \mathcal{J}^n} F_{J_j}^{rot}[i]$,
$\mathcal{J}^n = \{J_j: F_{J_j}^{road}=n\}$, and
$F_{J_j}^{rot}[i]=\alpha_i$ with $F_{J_j}^{rot}=[\alpha_1, \ldots, \alpha_n]$.

Hence, the \emph{map feature} of a set of maps is described as $\mathcal{F} = \bigcup_{n\in \mathcal{F}^{road}} \mathcal{F}_n^{rot} \times \mathcal{F}^{ctrl} \times \mathcal{F}^{xwlk}$.

\subsection{Map Construction Based on Feature Vectors}
\label{section:combinatorial_sampling}
During the map construction with the junctions generated by the feature vectors, we need to ensure 1) all junctions are connected, 2) the generated map emulates urban road networks, and 3) no road or junction overlaps, which makes the construction process challenging.
In this paper, we propose a greedy method to construct a grid-layout map iteratively. The details are shown in Algorithm \ref{alg:grid_layout_gen}.

\begin{algorithm}[t] \small
  \caption{Feature-based HD Map Generation}
  \label{alg:grid_layout_gen}
  
\KwIn{The set of junction features: $\mathcal{F} = \bigcup_{n\in \mathcal{F}^{road}} \mathcal{F}_n^{rot} \times \mathcal{F}^{ctrl} \times \mathcal{F}^{xwlk}$.}
\KwOut{A new map $\mathcal{M}_c$.}
  
  Initialize the map $\mathcal{M}_c=\emptyset$, the sets of empty grid points $\mathcal{P}_e=\{P_{0, 0}\}$ and filled grid points $\mathcal{P}_f=\emptyset$, an empty junction feature set $\mathcal{J}=\emptyset$; \label{alg:init}
  
  $j = 0$;
  
  \For{$n \in \mathcal{F}^{road}$}{ \label{alg:comb_start}
    \For{$control \in \mathcal{F}^{ctrl}$}{
        \For{$crosswalk \in \mathcal{F}^{xwlk}$}{
            $F^{road}_{J_j} = n$;
            
            $F_{J_j}^{rot} = [\alpha_i : \alpha_i \in \mathcal{U}_{[\mathcal{F}_{n}^{rot}[i][0],\mathcal{F}_{n}^{rot}[i][1]]}], \forall i \in \{1, ..., n\}$;
            
            $F_{J_j}^{ctrl} = control$;
            
            $F_{J_j}^{xwlk} = crosswalk$;
            
            $\mathcal{F}_{J_j} = [F^{road}_{J_j}, F_{J_j}^{rot}, F_{J_j}^{ctrl}, F_{J_j}^{xwlk}]$;\label{alg:comb_sample}
            
            $\mathcal{J} = \mathcal{J} \cup \{\mathcal{F}_{J_j}\}$, $j = j + 1$; \label{alg:comb_end}
        }
    }
  } 
    
  \While{$\mathcal{J} \neq \emptyset$}{
    Select a junction feature $\mathcal{F}_{J_j}$ and $\mathcal{J} = \mathcal{J} \setminus \{\mathcal{F}_{J_j}\}$; \label{alg:assign_complete}
    
    Find $\mathcal{F}_{J_j}$'s best match grid point in $\mathcal{P}_e$, say $P_{x,y}$; \label{alg:best_match}
    
    $\mathcal{M}_c=\mathcal{M}_c \cup \{(P_{x,y}, \mathcal{F}_{J_j})\}$; \label{alg:assignment}
    
    $\mathcal{P}_e = \mathcal{P}_e\setminus \{P_{x,y}\}$, 
    $\mathcal{P}_f = \mathcal{P}_f\cup \{P_{x,y}\}$;
    
    \If{$P_{x+1, y} \not\in \mathcal{P}_e \cup \mathcal{P}_f$}{ \label{alg:extend_start}
        $\mathcal{P}_e = \mathcal{P}_e \cup \{P_{x+1, y}\}$;
    }
    \If{$P_{x, y+1} \not\in \mathcal{P}_e \cup \mathcal{P}_f$}{
        $\mathcal{P}_e = \mathcal{P}_e \cup \{P_{x, y+1}\}$;
    }
    \If{$P_{x-1, y} \not\in \mathcal{P}_e \cup \mathcal{P}_f$}{
        $\mathcal{P}_e = \mathcal{P}_e \cup \{P_{x-1, y}\}$;
    }
    \If{$P_{x, y-1} \not\in \mathcal{P}_e \cup \mathcal{P}_f$}{
        $\mathcal{P}_e = \mathcal{P}_e \cup \{P_{x, y-1}\}$; \label{alg:extend_end}
    } 
  }
  \Return $\mathcal{M}_c$.
\end{algorithm}

Given a set of HD maps $\mathcal{M}$, by checking all junctions of the maps, we can extract the features $\mathcal{F}^{road}$, $\{\mathcal{F}_n^{rot}: n\in\mathcal{F}^{road}\}$, $\mathcal{F}^{ctrl}$, and $\mathcal{F}^{xwlk}$.
Before constructing the map, we initialize a grid point container $\mathcal{P}_e = \{P_{0,0}\}$ for holding empty points, another grid point container $\mathcal{P}_f = \emptyset$ for filled points and a set $\mathcal{J} = \emptyset$ for storing sampled junction features~(Line \ref{alg:init}).
For each vector $[F^{road}_{J_j}, F_{J_j}^{ctrl}, F_{J_j}^{xwlk}]$ in $\mathcal{F}^{road} \times \mathcal{F}^{ctrl} \times \mathcal{F}^{xwlk}$, we perform multivariate uniform sampling in the region $\prod_{i=1}^n [\mathcal{F}_{n}^{rot}[i][0],\mathcal{F}_{n}^{rot}[i][1]]$ to generate a junction feature  $\mathcal{F}_{J_j} = [F^{road}_{J_j}, F_{J_j}^{rot}, F_{J_j}^{ctrl}, F_{J_j}^{xwlk}]$~(Lines \ref{alg:comb_start} - \ref{alg:comb_sample}).
The sampled junction features are stored in $\mathcal{J}$ (Line  \ref{alg:comb_end}), each of which will be used to instantiate an individual junction in the grid layout.

Next, for each junction feature $\mathcal{J}_j$, we search for the empty grid points in $\mathcal{P}_e$ to select the grid point $P_{x,y}$ of the highest matching score with $\mathcal{J}_j$. The matching score is heuristically calculated in the following way: 

Let $k$ be the maximal number of roads that $J_j$ at $P_{x,y}$ can connect after proper rotation with the surrounding junctions (if available) at $P_{x+1, y}$, $P_{x, y+1}$, $P_{x-1,y}$, and $P_{x, y-1}$. If $F^{road}_{J_j} > k$, then the feature and the point does not match and the score is set to $-\infty$. Otherwise, the score is set to $k$. Random selection is performed when a draw occurs. 

After the grid point, say $P_{x,y}$, is selected for $\mathcal{F}_{J_j}$, and the corresponding junction is instantiated at $P_{x,y}$~(Line \ref{alg:best_match}-\ref{alg:assignment}).
The junction feature is removed from the feature set (Line \ref{alg:assign_complete}) and the grid is extended in the four regular directions (i.e., \texttt{East}, \texttt{North}, \texttt{West}, and \texttt{South}) from $P_{x,y}$, free for new junction assignments (Line \ref{alg:extend_start}-\ref{alg:extend_end}).
After all the junctions are instantiated, the roads connecting junctions at neighboring grid points can be constructed via road socket vectors (controlled by $F^{rot}$). The \emph{start point} (resp. \emph{end point}) of road lanes can then be obtained by shifting the \emph{start point} (resp. \emph{end point}) of the road's reference path along the perpendicular direction of the \emph{start heading} (resp. \emph{end heading}) of the road. After the road lanes are built, the junction lanes can be added by connecting the incoming and outgoing lanes of the junction. Next, traffic controls and crosswalks are created at appropriate locations. 
Lastly, we can output the HD map files based on $\mathcal{M}_c$.

Note that to make the generated HD map as concise as possible while preserving potential scenario diversity, we set the following default configurations for \tool:
\begin{itemize}[leftmargin=*]
    \item Given two junctions with the same junction feature, but one contains a one-way road while the other contains only two-way roads, any testing scenario generated from the former can also be generated from the latter one. Hence, all roads are two-way to keep the generated map concise. 
    \item To minimize the number of junctions in the generated map, we assume that all roads are fully connected at junctions. It means there exists at least one navigation path (i.e., junction lane) connecting each road pair.
    \item The crosswalks are instantiated within the range of junctions and only overlap with junction lanes instead of road lanes. The signals are placed in front of the road it controls (facing the road), and the stop signs are placed on the right side of the road (facing the road). 
    \item The default grid-gap is 100 meters (e.g., $P_{x+1, y}$ is 100 meters to the east of $P_{x,y}$, lane width is 3.5 meters (default value of selected input maps), and crosswalk width is 4.5 meters (slightly longer than a regular sedan).
\end{itemize}

\section{Experiments}
\label{section:experiments}


\begin{table}[]
\centering
\caption{Features extracted from San Francisco map}
\label{tab:feature-result}
\begin{tabular}{>{\centering}m{0.05\linewidth}cccc}
\hline
$\mathcal{F}^{road}$ & $\mathcal{F}^{ctrl}$ & $\mathcal{F}^{xwlk}$ & $\mathcal{F}^{rot}_3$ & $\mathcal{F}^{rot}_4$ \\ \hline
\{3, 4\} & \begin{tabular}[c]{@{}c@{}}$\{signal$, \\ $stop\}$\end{tabular} & \begin{tabular}[c]{@{}c@{}}$\{True$, \\$ False\}$\end{tabular} & \begin{tabular}[c]{@{}c@{}}{[}{[}-22.62,9.81{]},\\ {[}61.40, 126.22{]},\\ {[}-143.45, 200.92{]}{]}\end{tabular} & \begin{tabular}[c]{@{}c@{}}{[}{[}-5.26, 13.61{]},\\ {[}-75.63, 93.10{]},\\ {[}-176.85, 191.76{]},\\ {[}-109.24, -87.91{]}{]}\end{tabular} \\ \hline
\end{tabular}
\end{table}

\begin{figure}
    \centering
    \subfigure[Input map.]{
        \begin{minipage}[b]{0.2\columnwidth}
            \centering
            \label{fig:san_map} 
            \includegraphics[width=2cm]{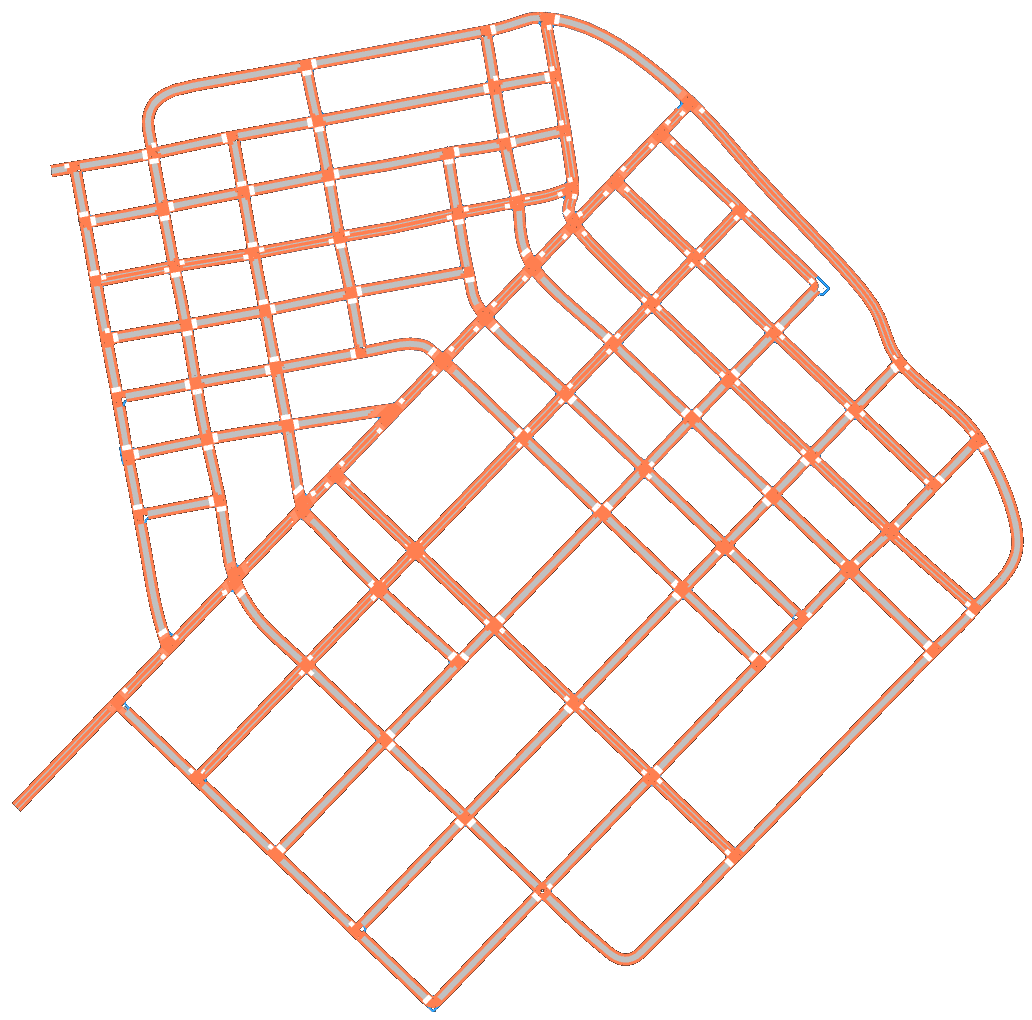}
        \end{minipage}} 
    \hspace{2pt}
    \subfigure[$M_1$]{
        \label{fig:feature_4_map}
        \begin{minipage}[b]{0.15\columnwidth}
            \centering
            \label{fig:gen_map_1} 
            \includegraphics[width=1.3cm]{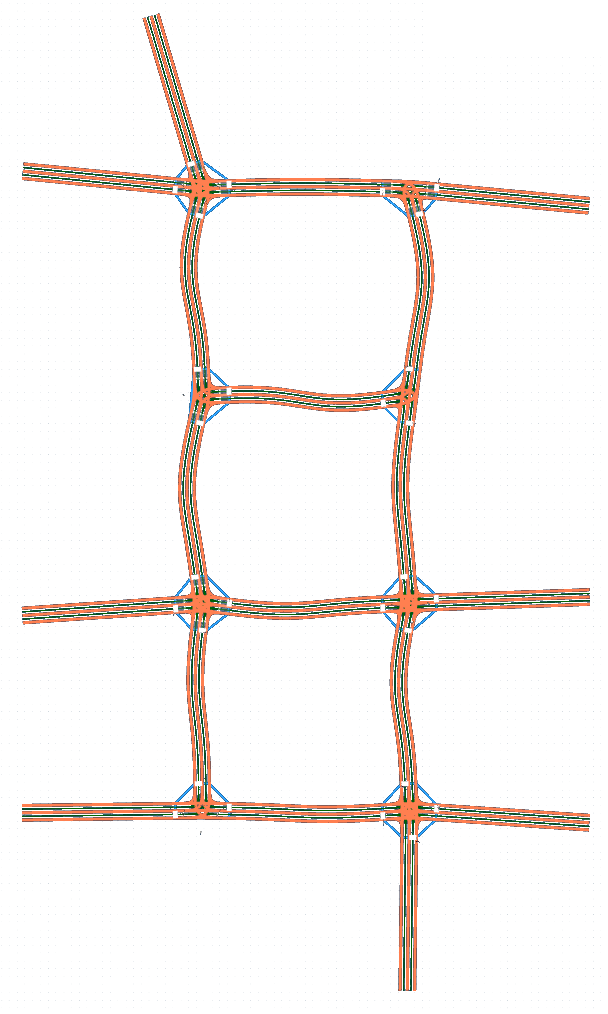}
        \end{minipage}} 
    \hspace{2pt}
    \subfigure[$M_2$]{
        \begin{minipage}[b]{0.2\columnwidth}
            \centering
            \label{fig:gen_map_2} 
            \includegraphics[width=2cm]{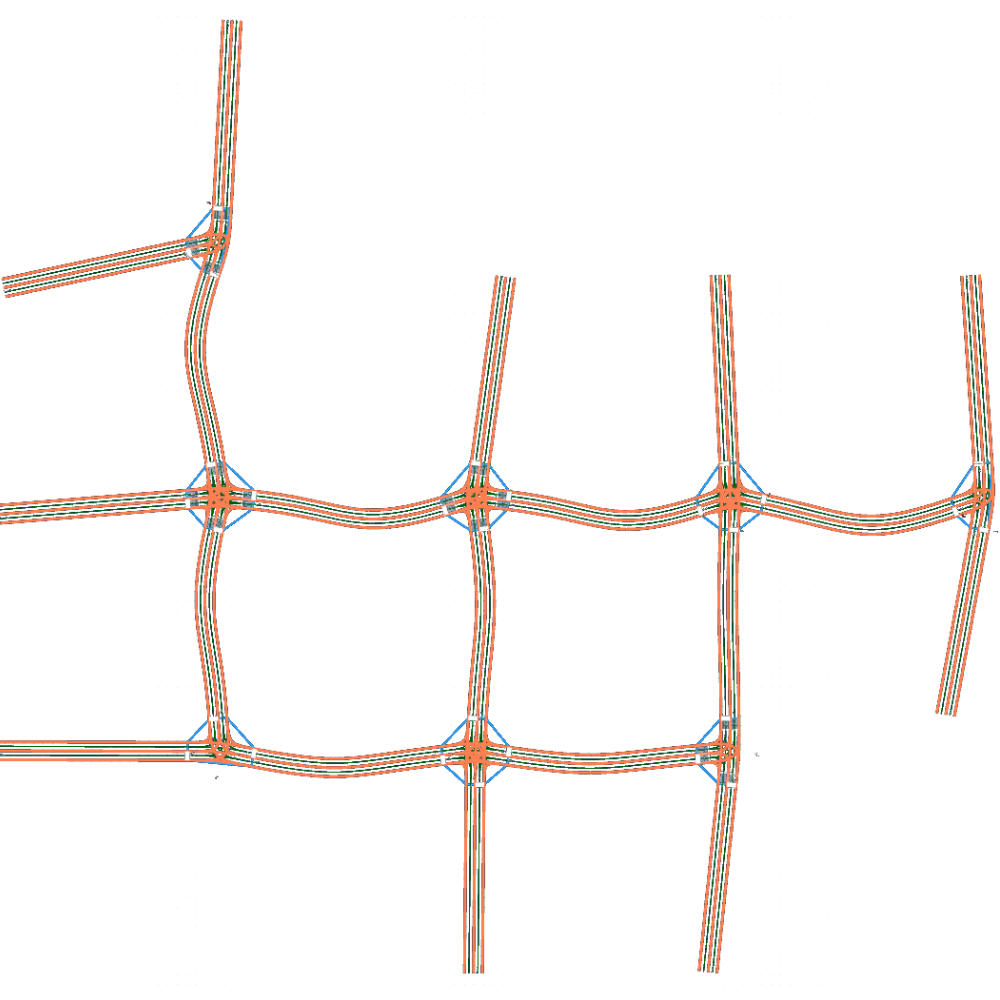}
        \end{minipage}} 
    \hspace{2pt}
    \subfigure[$M_3$]{
        \begin{minipage}[b]{0.2\columnwidth}
            \centering
            \label{fig:gen_map_3} 
            \includegraphics[width=2cm]{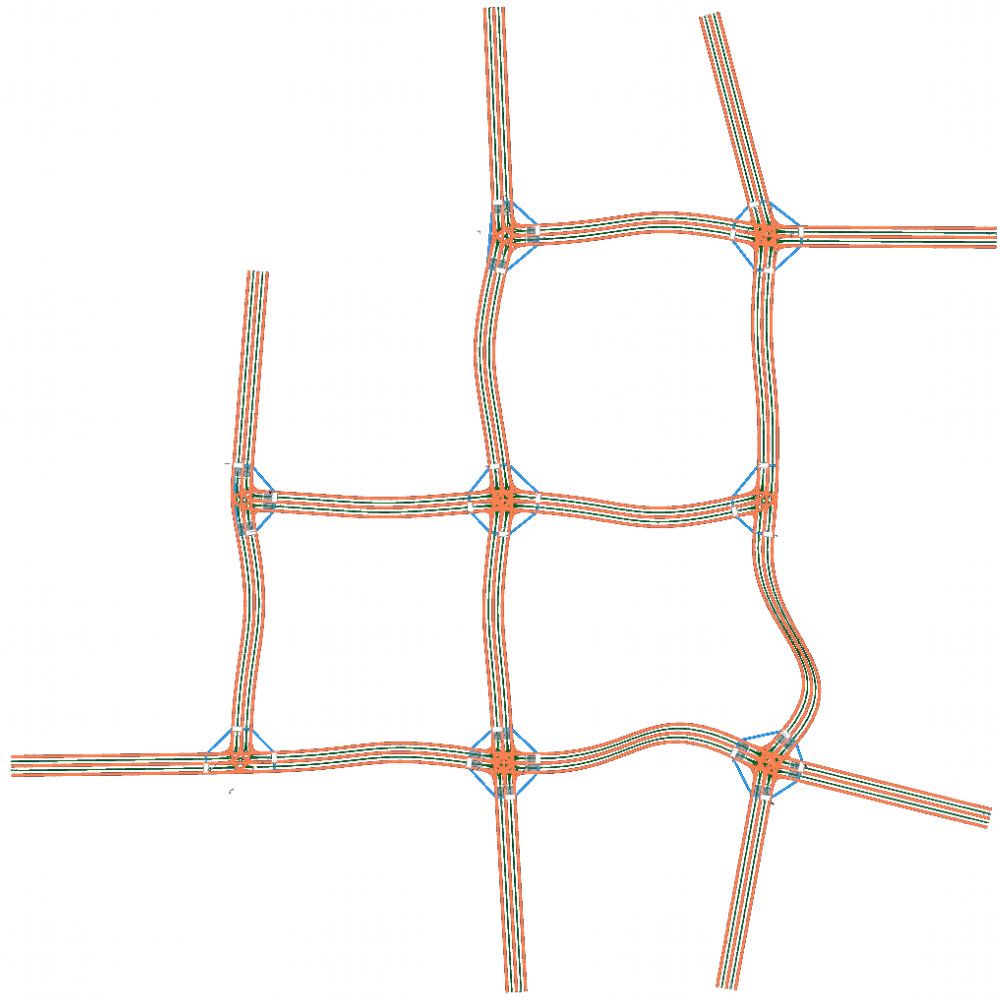}
        \end{minipage}}
    \caption{(a) The input $San Francisco$ map; (b)--(d) three generated HD maps $M_1$, $M_2$, and $M_3$.}
    \label{fig:maps}
\end{figure}

We conduct comprehensive experiments in this section to answer the following research questions.

\noindent \textbf{RQ1}: Can \tool generate complete HD maps connecting all the intersections from sampled features correctly?

\noindent \textbf{RQ2}: Can \tool take a single input map and generate concise maps of the same level of scenario diversity as the input?

\noindent \textbf{RQ3}: Can \tool take multiple input maps and generate concise maps of equal or greater scenario diversity than individual input?

\noindent \textbf{RQ4}: Can \tool take user-specified features and generate customized maps?

\begin{figure}
    \centering
    \subfigure[$J_0$]{
        \begin{minipage}[b]{0.22\columnwidth}
            \centering
            \label{fig:gen_junction_1} 
            \includegraphics[width=2.1cm]{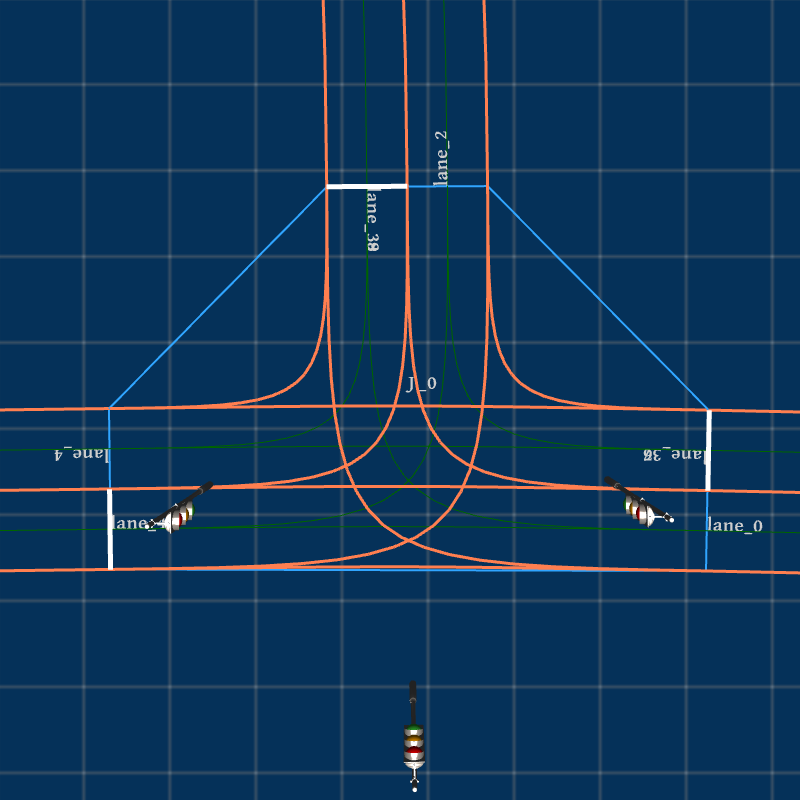}
        \end{minipage}} 
    \subfigure[$J_1$]{
        \begin{minipage}[b]{0.22\columnwidth}
            \centering
            \label{fig:gen_junction_2} 
            \includegraphics[width=2.1cm]{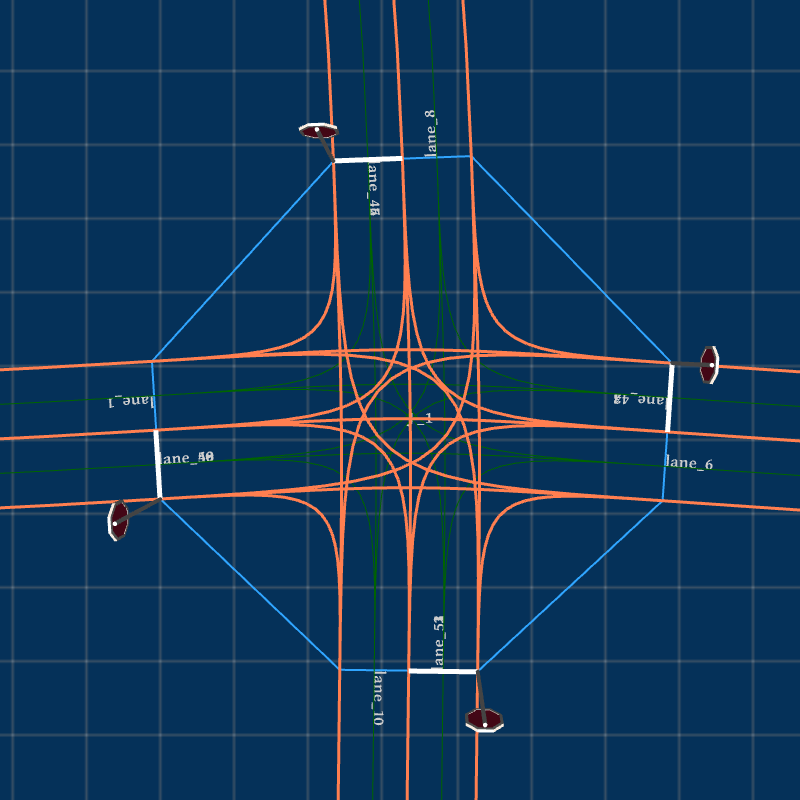}
        \end{minipage}} 
    \subfigure[$J_2$]{
        \begin{minipage}[b]{0.22\columnwidth}
            \centering
            \label{fig:gen_junction_3} 
            \includegraphics[width=2.1cm]{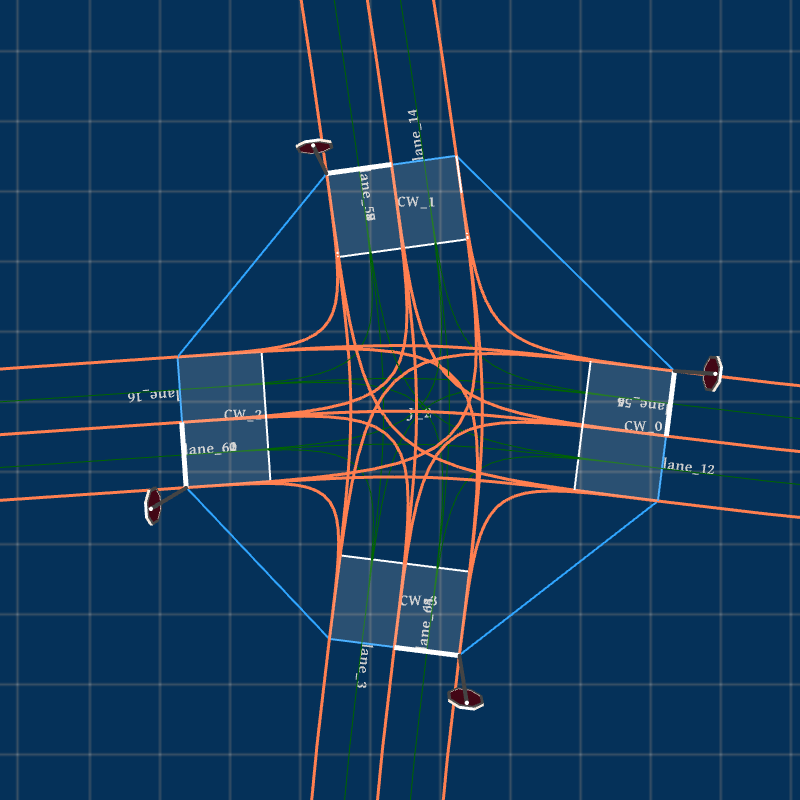}
        \end{minipage}} 
    \subfigure[$J_3$]{
        \begin{minipage}[b]{0.22\columnwidth}
            \centering
            \label{fig:gen_junction_4} 
            \includegraphics[width=2.1cm]{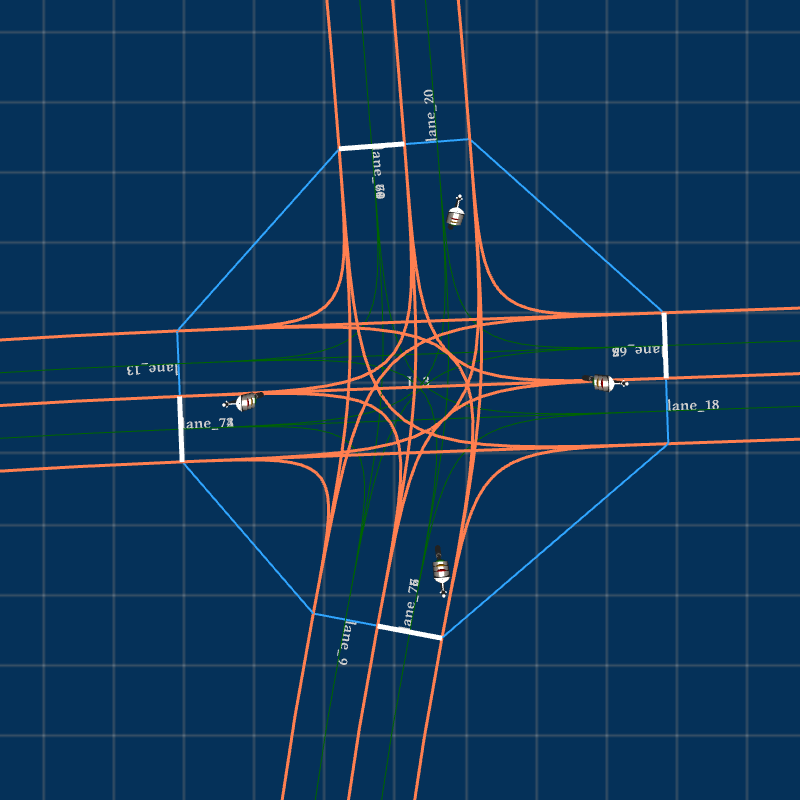}
        \end{minipage}}
    \subfigure[$J_4$]{
        \begin{minipage}[b]{0.22\columnwidth}
            \centering
            \label{fig:gen_junction_5} 
            \includegraphics[width=2.1cm]{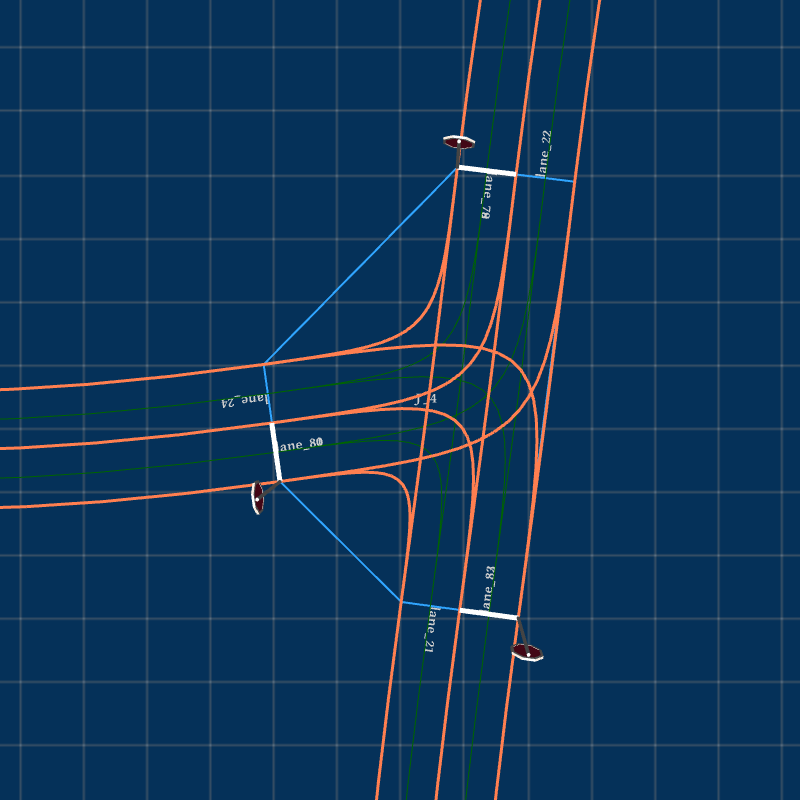}
        \end{minipage}}
    \subfigure[$J_5$]{
        \begin{minipage}[b]{0.22\columnwidth}
            \centering
            \label{fig:gen_junction_6} 
            \includegraphics[width=2.1cm]{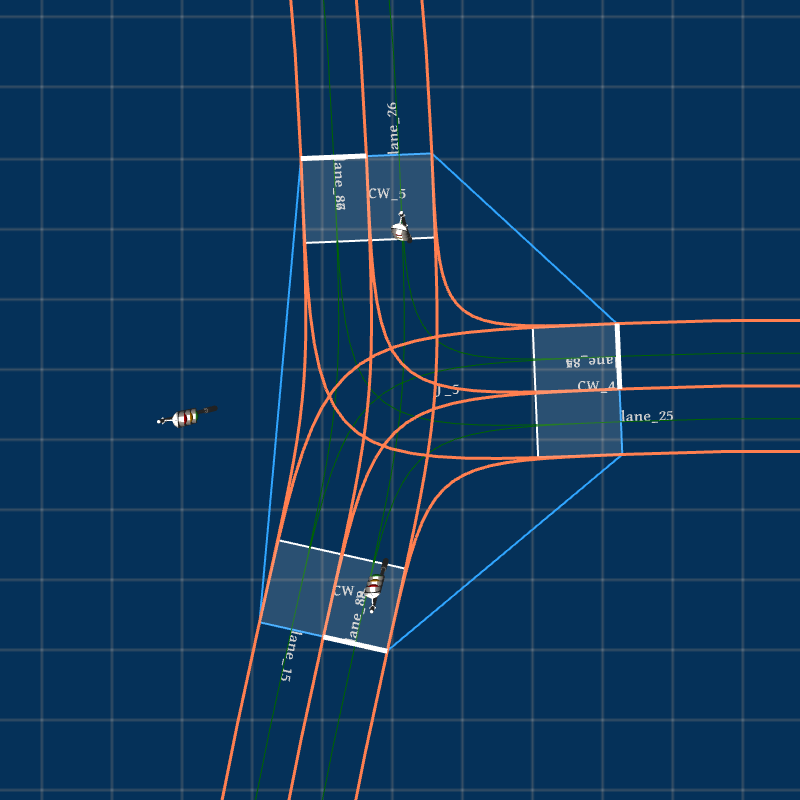}
        \end{minipage}}
    \subfigure[$J_6$]{
        \begin{minipage}[b]{0.22\columnwidth}
            \centering
            \label{fig:gen_junction_7} 
            \includegraphics[width=2.1cm]{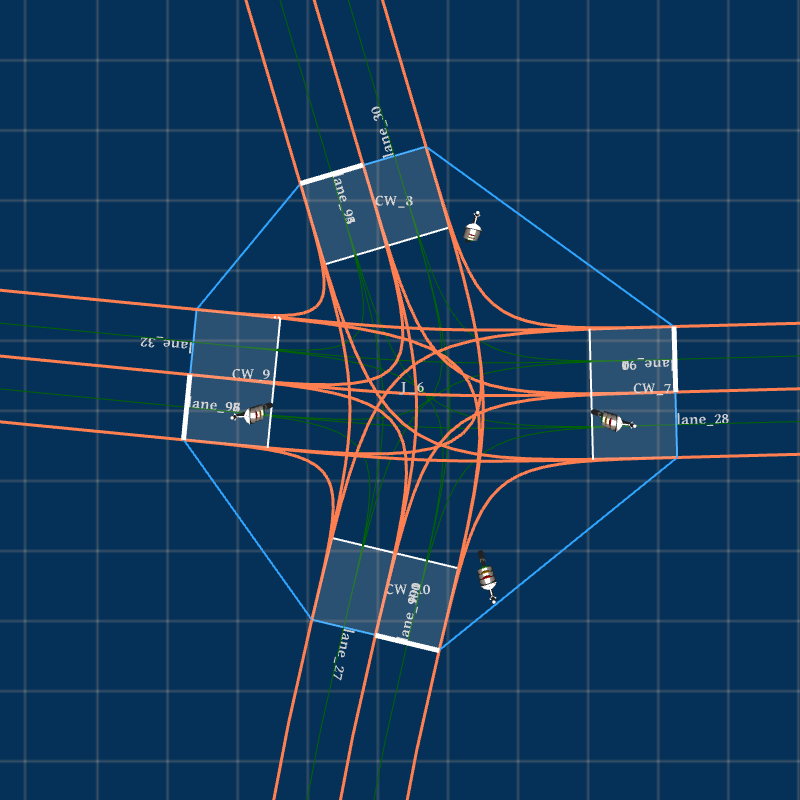}
        \end{minipage}}
    \subfigure[$J_7$]{
        \begin{minipage}[b]{0.22\columnwidth}
            \centering
            \label{fig:gen_junction_8} 
            \includegraphics[width=2.1cm]{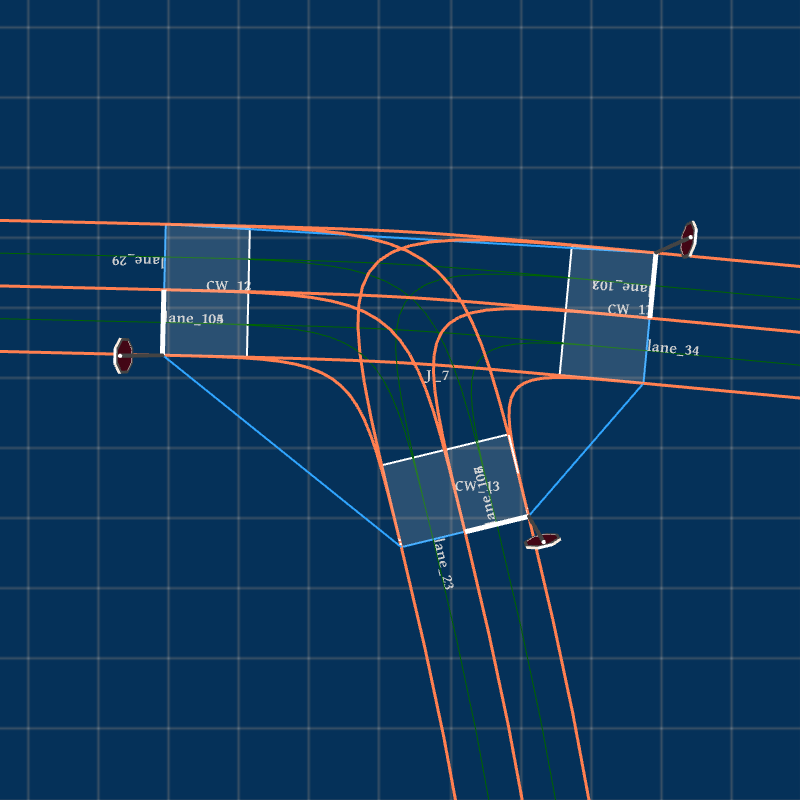}
        \end{minipage}}
    \caption{The detailed junctions of the map in Fig. \ref{fig:feature_4_map}.}
    \label{fig:generated_junctions}
\end{figure}

\subsection{Experimental Settings}

To demonstrate the effectiveness of \tool, we implement a prototype of \tool 
and conduct experiments on the latest Apollo 7.0 \cite{web_apollo_github} stack with three of the largest HD maps available for Apollo, i.e., $San Francisco$ \cite{web_san_map}, $Go Mentum$ \cite{web_gomentum_map} and $Shalun$ \cite{web_shalun_map}. We define two metrics to measure scenario diversity.
The first one is \emph{line coverage}, which describes how many statements in the ADS have been executed by the scenarios. Two scenarios are considered different if they cover different statements. 
The other one is \emph{behavior coverage}, which describes different behavior transitions during the execution of a scenario. 

To answer RQ1, we run \tool on $San Francisco$ and perform qualitative analysis on three generated maps $M_{1-3}$.
To answer RQ2, we conduct quantitative analysis by comparing the coverage results on $San Francisco$ and $M_{1-3}$.
To answer RQ3, we generate a new map $M_4$ from $San Francisco$ \cite{web_san_map}, $Go Mentum$ \cite{web_gomentum_map} and $Shalun$ \cite{web_shalun_map} and conduct the same experiments as in RQ2.
To answer RQ4, we illustrate the construction of customized maps from manual features.

All the experiments are conducted on a desktop computer (with i7-6850K CPU, 64GB RAM, and GTX 1080T Ti). When an HD map is generated, \emph{Dreamview} (Apollo's full-stack HMI service) \cite{web_dreamview} is restarted to load the new map. We run the route coverage testing \cite{tang2021route} in Apollo's built-in \emph{sim-control} mode \cite{web_sim_control}, which does not require any third-party simulators such as SVL \cite{svl_simulator}. The input of the planning component is published using simulated dummy data. Specifically, \emph{sim-control} publishes the results of $localization$, $chassis$, and $prediction$, while our testing script submits $routing\ requests$ and publishes $traffic\ light$ information at runtime.

\subsection{RQ1: Qualitative Analysis on Map Generation}

\emph{San Francisco} map contains 91 junctions (including 88 signal-controlled and three stop-sign controlled), 164 roads, and 1524 lanes (including 868 junction lanes and 656 road lanes); there are 13 junctions with crosswalks. 

Due to randomness in the rotation sampling of road socket vectors, the extracted feature set $\mathcal{F}$ can generate unlimited HD maps.
Fig. \ref{fig:maps} presents the bird-view of the input map $San Francisco$ and three generated HD maps $M_1$, $M_2$ and $M_3$ based on extracted features listed in Table \ref{tab:feature-result}. 
We can find that the number of junctions is reduced to 8 in $M_{1-3}$ from 91 in $San Francisco$.

Take $M_1$ in Fig. \ref{fig:feature_4_map} for example. 
According to Algorithm \ref{alg:grid_layout_gen} given in Section  \ref{section:combinatorial_sampling},
\tool generates 8 junctions, as shown in Fig. \ref{fig:generated_junctions}. Junction $J_0$, $J_4$, $J_5$, and $J_7$ are three-legged T-junctions and the rest are four-legged intersections. Junction $J_0$, $J_3$, $J_5$, and $J_6$ are signal-controlled while the rest are stop-sign controlled. Junction $J_2$, $J_5$, $J_6$, and $J_7$ have crosswalks instantiated at all the connected roads while the rest have no crosswalks. From Fig. \ref{fig:generated_junctions}, we can see that \tool can generate junctions with different topological and geometrical features and instantiate traffic controls and crosswalks correctly. 

\begin{figure}[t]
    \centering
    \subfigure[Code coverage.]{
        \begin{minipage}[b]{\columnwidth}
            \centering
            \label{fig:comparison_code} 
            \includegraphics[width=8.5cm]{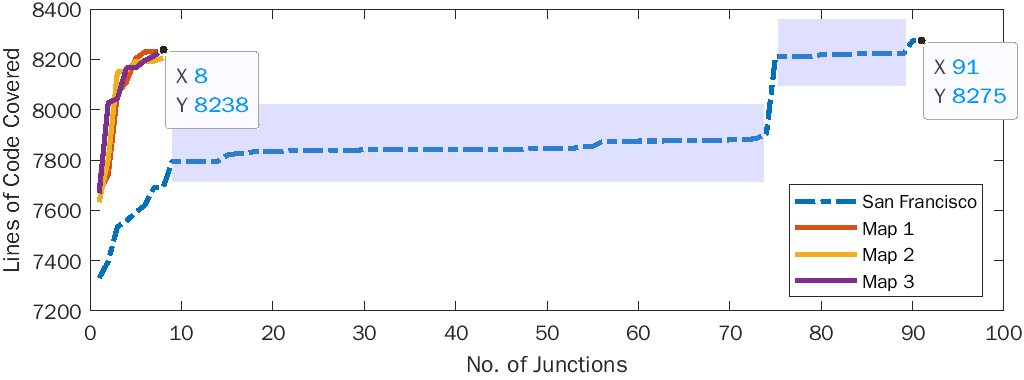}
        \end{minipage}}
    \subfigure[Stage transition.]{
        \begin{minipage}[b]{\columnwidth}
            \centering
            \label{fig:comparison_stage} 
            \includegraphics[width=8.5cm]{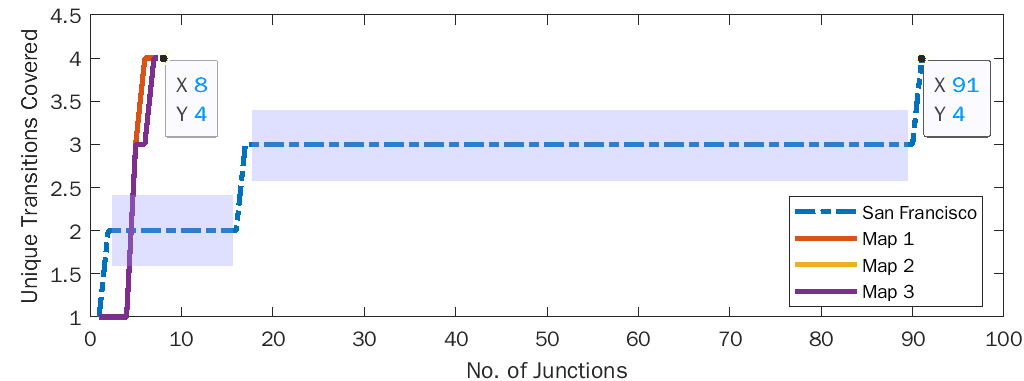}
        \end{minipage}} 
    \caption{Comparison of code coverage and stage transitions of Apollo running on $San Francisco$ map and the three generated maps.}
    \label{fig:comparison}
\end{figure}

\subsection{RQ2: Quantitative Analysis of Scenario Diversity}


To collect the line coverage, we instrument Apollo's planning component during compilation using the \emph{coverage} command \cite{web_bazel_coverage} provided by its build tool \emph{Bazel} \cite{web_bazel} as the scenario handling mechanisms are implemented in the planning component.
Fig. \ref{fig:comparison} shows the coverage results of our experiments on the four maps in Fig. \ref{fig:maps}.
Fig. \ref{fig:comparison_code} shows the accumulated code coverage results on different maps. The accumulated lines covered eventually are 8275 ($San Francisco$), 8232 ($M_1$), 8205 ($M_2$) and 8238 ($M_3$) respectively. 
As we can see, $San Francisco$ has many junctions covering the same lines of code (highlighted in light blue), resulting in a significant amount of duplicated test cases. On the other hand, unique lines of code get covered in every junction in $M_{1-3}$. It only takes eight junctions in $M_{1-3}$ to reach a similar level of code coverage. Notice that $San Francisco$ has slightly higher code coverage than $M_{1-3}$. Based on the source code-level investigation, the uniquely covered codes in $San Francisco$ are related to human errors in the HD map file (e.g., the id of some crosswalk elements are missing). Such human errors do not exist in maps generated by \tool automatically. Besides the map error, we also discovered some potential issues of the planning component (reported at \cite{web_apollo_issue_1, web_apollo_issue_2}). Such issues, encountered in different situations (e.g., on roads or inside junctions), trigger different error handling mechanisms implemented in the planning component. This is the main reason for the slight difference in the line coverage results of $M_{1-3}$. 

\begin{figure}
    \centering
    \includegraphics[width=\linewidth]{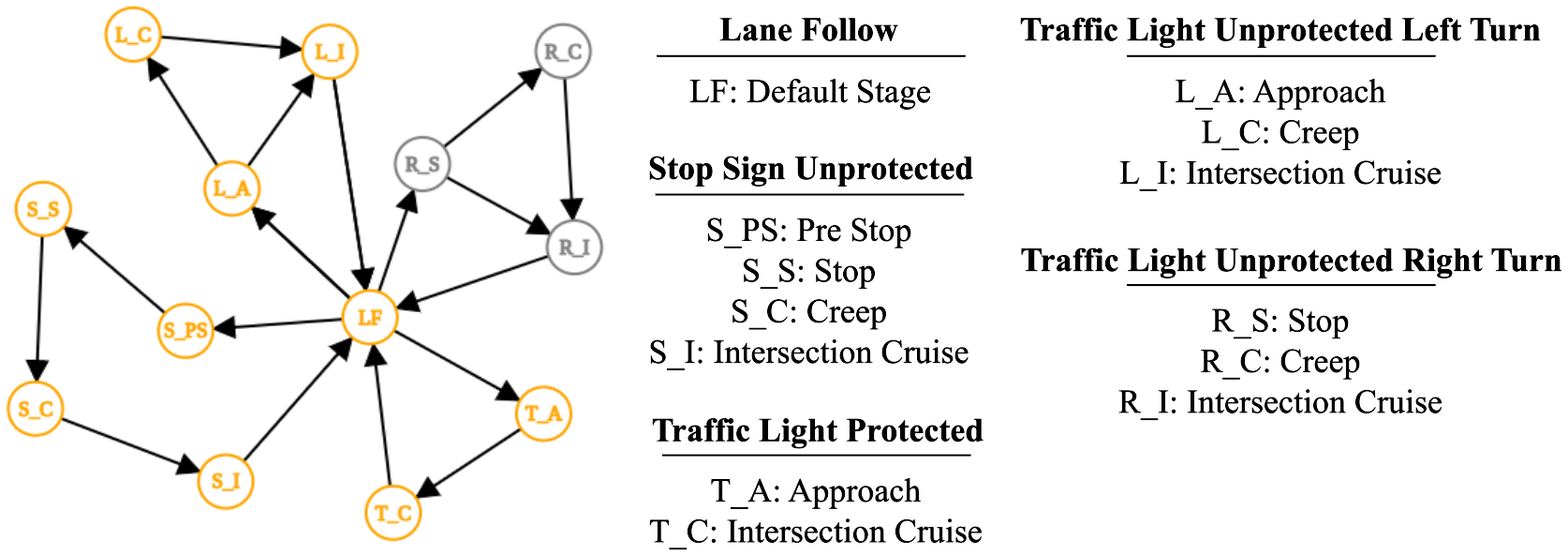}
    \caption{Scenarios and stage transitions covered in $San Francisco$ and $M_{1-3}$.}
    \label{fig:stage_transitions} 
\end{figure}

Apollo implements its planning component as a modularized state machine, where states align with common traffic scenarios such as intersection cruising, emergency stops, etc. and publishes its internal scenario selections (e.g., \emph{LaneFollow} and \emph{TrafficLightProtected}) and stage transitions (e.g. \emph{Approach} and \emph{IntersectionCruise}) within each scenario. 

The coverage results of the stage transitions are shown in Fig. \ref{fig:comparison_stage}. From the results, we can find that many junctions belong to the same scenario in Apollo's view of point. As a result, the same strategies are applied by Apollo to navigate through them (highlighted in light blue as well). On the other hand, the eight junctions in $M_{1-3}$ reach the same level of coverage as that of 91 junctions in the $San Francisco$ map. Specifically, there are 16 scenario types defined in Apollo \cite{web_apollo_planning_config_proto}, out of which five scenarios and their stages are relevant to the experiments \cite{web_apollo_scenario_manager_code}: 
\begin{itemize}
    \item \emph{Lane Follow}
    \item \emph{Stop Sign Unprotected}
    \item \emph{Traffic Light Protected}
    \item \emph{Traffic Light Unprotected Left Turn}
    \item \emph{Traffic Light Unprotected Right Turn}
\end{itemize}

The relevant scenarios and their stage transitions are plotted in Fig. \ref{fig:stage_transitions}, where the nodes represent the stages and arrows are possible transitions. The stages and transitions covered by $San Francisco$ and $M_{1-3}$ are the same and highlighted in orange, while the uncovered stages are rendered grey. 

The four paths covered are:

\begin{itemize}
    \item $LF \rightarrow S\_PS \rightarrow S\_S \rightarrow S\_C \rightarrow S\_I \rightarrow LF$
    \item $LF \rightarrow T\_A \rightarrow T\_C \rightarrow LF$
    \item $LF \rightarrow L\_A \rightarrow L\_I \rightarrow LF$
    \item $LF \rightarrow L\_A \rightarrow L\_C \rightarrow L\_I \rightarrow LF$
\end{itemize}

The stage transitions align with our testing setup, starting and ending at road lanes (i.e., $LF$). The covered scenarios also match with the features of the generated junctions. One may notice that the right turn-related scenarios are not covered. This is because Apollo only enters the right turn scenario when the ADS is turning right under the red traffic light. When the traffic light is green, which is always the case during our $sim-control$ simulation, Apollo treats right turning the same as driving straight across the junction (i.e., $T\_A$ and $T\_C$).

\subsection{RQ3: Map Generation From Multiple Maps}

We generate map $M_4$ with multiple input maps, i.e., $San Francisco$, $Go Mentum$, and $Shalun$.
The experiment results are listed in Table \ref{tab:multi_input_map_gen}. The features extracted from $Shalun$ are not much different from $San Francisco$. However, due to the existence of the bare junctions (i.e., no controls) in $Go Mentum$, the merged control feature is $\mathcal{F}^{ctrl}=\{signal, stop, bare\}$, which results in 12 junctions to be generated in $M_4$. 
The code coverage of $M_4$ is similar to those of $M_{1-3}$.
We find that Apollo 7.0 treats bare junctions as normal roads, hence no significant difference in the code coverage. 
However, a new path consisting of only one stage $LF$ (i.e., no junction scenarios are triggered) is covered because of the existence of bare junctions. 

In summary, experiments with single and multiple input maps have demonstrated the effectiveness of \tool in feature-based map generation.

\begin{figure}
    \centering
    \subfigure[Mega junction]{
        \begin{minipage}[b]{0.295\columnwidth}
            \centering
            \label{fig:manual_junction_1} 
            \includegraphics[width=2.75cm]{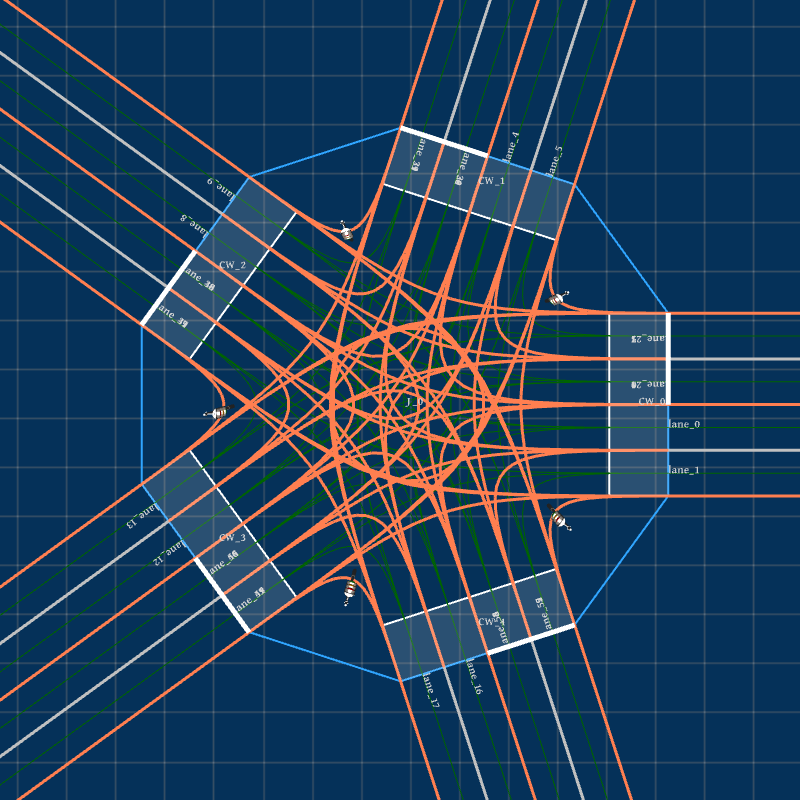}
        \end{minipage}} 
    \subfigure[Road circle]{
        \begin{minipage}[b]{0.295\columnwidth}
            \centering
            \label{fig:manual_road} 
            \includegraphics[width=2.75cm]{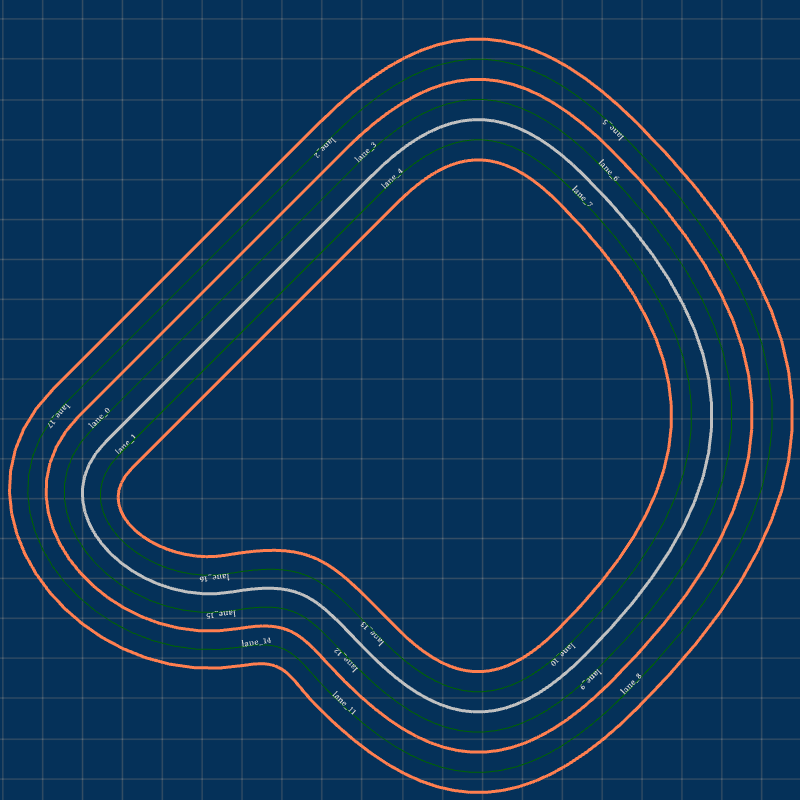}
        \end{minipage}} 
    \subfigure[One-way roads]{
        \begin{minipage}[b]{0.295\columnwidth}
            \centering
            \label{fig:manual_junction_2}
            \includegraphics[width=2.75cm]{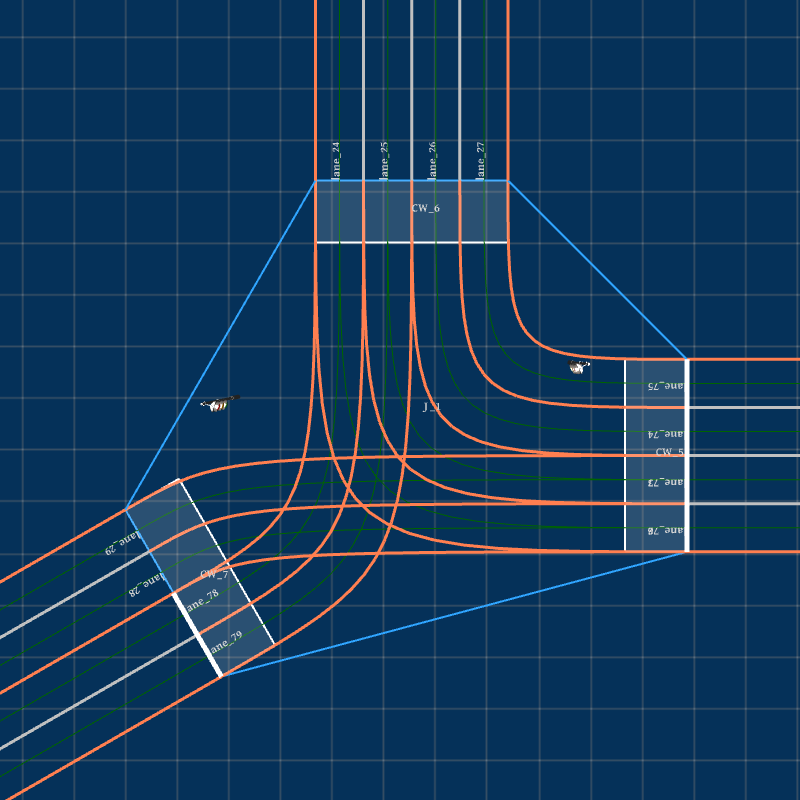}
        \end{minipage}} 
    \caption{Junctions generated with manually defined junction features.}
    \label{fig:manual_maps}
\end{figure}

\begin{table}
\centering
\caption{Multiple input map generation}
\label{tab:multi_input_map_gen}
\begin{tabular}{cccc}
\hline
Discrete Features & Rotation Features & Code & Stage \\ \hline
\begin{tabular}[c]{@{}c@{}}$\mathcal{F}^{road} = \{3, 4\}$\\ \\ $\mathcal{F}^{ctrl} =$ \\ $\{signal, stop, bare\}$\\ \\ $\mathcal{F}^{xwlk} = \{True, False\}$\end{tabular} & \begin{tabular}[c]{@{}c@{}}$\mathcal{F}^{rot}_3 =$\\ {[}{[}-23.58, 28.03{]},\\ {[}61.4, 127.39{]},\\ {[}140.41, 213.36{]}{]}\\ \\$\mathcal{F}^{rot}_4 =$\\ {[}{[}-7.49, 13.61{]},\\ {[}60.46, 134.06{]},\\ {[}142.18, 214.47{]},\\ {[}-109.24, -69.95{]}{]}\end{tabular} & 8236 & \begin{tabular}[c]{@{}c@{}}4 paths \\ covered \\ in previous\\  section\\ + \\ $LF$\end{tabular} \\ \hline
\end{tabular}
\end{table}

\subsection{RQ4: Map Generation from User-Defined Features}

The generation capability of \tool is not limited to generating junctions from the existing maps, such as those shown in Fig. \ref{fig:generated_junctions}. 
\tool also accepts manually defined feature values if users require customization.
For example, the mega junction shown in Fig. \ref{fig:manual_junction_1} is generated by specifying the following junction feature: $F_{mega}^{road} = 5$, $F_{mega}^{ctrl}=signal$, $F_{mega}^{xwlk}=True$ and $F_{mega}^{rot}=[0^\circ, 72^\circ, 144^\circ, -144^\circ, -72^\circ]$.

With the cubic Bézier curve model, \tool can also generate a set of roads connected end-to-end by specifying a list of road configurations where each road's \emph{end point} (resp. \emph{end heading}) is the next road's \emph{start point} (resp. \emph{start heading}).
Finally, in addition to the default two-way roads, one-way roads are also supported by \tool as long as users specify the road type of each road socket. Fig. \ref{fig:manual_junction_2} shows a T-junction where the road types associated to the sequence of the road sockets, started from the \texttt{East} most road socket and listed in a counterclockwise direction, is $[In, Out, In Out]$, where $In$, $Out$, $InOut$ mean one-way incoming, one-way outgoing, and two-way roads, respectively. 

\section{Discussion and Conclusion}
\label{section:conclusion}


\noindent \textbf{Discussion.}
Scenario-based testing is inevitable across the entire phase of ADS development. How diverse and representative the testing scenarios are essentially determines the testing efficiency and effectiveness. 
Bagschik et al. \cite{bagschik2018ontology} proposes a 5-layer model for testing scenarios, including \emph{road level} (layer 1), \emph{traffic infrastructure} (layer 2), \emph{manipulation of layer 1 and 2} (layer 3), \emph{object} (layer 4) and \emph{environment} (layer 5). A recent survey \cite{zhong2021survey} shows that most works focus on sampling critical scenarios at layer 4 and layer 5, i.e., by controlling the trajectories of other traffic participants such as vehicles or pedestrians or manipulating the simulated weather. However, the topological and geometric features of the traffic networks (i.e., layer 1-3), which we believe are the foundation of testing scenarios, are often overlooked.
As a result, HD map generation based on required features substantially lifts the restrictions induced by the fixed elements of the first three layers and creates huge potential in generating diverse scenarios. 

\noindent \textbf{Conclusion.} 
In this paper, we propose an automatic feature-based HD map generation framework \tool for simulation-based testing. The framework takes in existing maps as input, extracts essential features, and applies combinatorial sampling to generate grid-layout HD maps. \tool also accepts manual feature configurations to allow flexible customization. Experiment results show that the generated HD map is concise and preserves comparable scenario diversity.

The topological and geometrical features of the roads and junctions can be far more complex in real life than those considered here. Our framework can be extended to more complex features such as filter lanes in mega junctions, crosswalks on road lanes, and highway traffic networks. We leave the exploration of them for future work.


\bibliographystyle{IEEEtran}
\bibliography{reference}
\end{document}